\documentclass[reprint,aps, prl, amsmath, amssymb, superscriptaddress, longbibliography, floatfix]{revtex4-1}
\usepackage{amsmath}
\usepackage{bm}
\usepackage[normalem]{ulem}
\usepackage{graphicx}
\usepackage{braket}
\usepackage{float}
\restylefloat{table}
\usepackage[colorlinks, linkcolor= blue, citecolor = blue, urlcolor=blue]{hyperref}

\def\be{\begin{equation}}
\def\ee{\end{equation}}
\def \bea{\begin{eqnarray}}
\def \eea{\end{eqnarray}}

\begin{document}
\title{Intrinsic nonlinear conductivities induced by the quantum metric}
\author{Kamal Das}
\email{daskamal457@gmail.com}
\affiliation{Department of Physics, Indian Institute of Technology Kanpur, Kanpur 208016, India}
\affiliation{Department of Condensed Matter Physics, Weizmann Institute of Science, Rehovot 7610001, Israel}
\thanks{K. D. and S. L. contributed equally to this work.}
\author{Shibalik Lahiri}
\email{lahiris@purdue.edu}
\affiliation{Department of Physics, Indian Institute of Technology Kanpur, Kanpur 208016, India}
\affiliation{ Department of Physics and Astronomy, Purdue University, West Lafayette, Indiana 47907, USA}
\thanks{K. D. and S. L. contributed equally to this work.}
\author{Rhonald Burgos Atencia}
\email{r.burgos@unsw.edu.au}
\affiliation{School of Physics, The University of New South Wales, Sydney 2052, Australia}
\affiliation{Australian Research Council Centre of Excellence in Low-Energy Electronics Technologies, The University of New South Wales, Sydney 2052, Australia}
\author{Dimitrie Culcer}
\email{d.culcer@unsw.edu.au}
\affiliation{School of Physics, The University of New South Wales, Sydney 2052, Australia}
\affiliation{Australian Research Council Centre of Excellence in Low-Energy Electronics Technologies, The University of New South Wales, Sydney 2052, Australia}
\author{Amit Agarwal}
\email{amitag@iitk.ac.in}
\affiliation{Department of Physics, Indian Institute of Technology Kanpur, Kanpur 208016, India}

\begin{abstract}
The second-order nonlinear current originates from three physical mechanisms: extrinsic nonlinear Drude and Berry curvature dipole and intrinsic Berry connection polarizability. Here, we predict a new intrinsic contribution to the current related to the quantum metric, a quantum geometric property of the electronic wave function. This contribution manifests in systems that simultaneously break the time-reversal and the inversion symmetry. {Interestingly, the new contribution is dissipative and contributes to both longitudinal and the dissipative nonlinear Hall response.} The quantum metric-induced NL current dominates transport in parity-time reversal symmetric systems near the band edges, something we show explicitly for topological antiferromagnets.
\end{abstract}

\maketitle


{\it Introduction:---} 
The nonlinear (NL) conductivity provides new physical insight into the quantum geometry of the electronic wave-function~\cite{moore_PRL2010_confin,deyo_arxiv2009_semi,sodemann_PRL2015_quantum,nandy_PRB2019_symmetry,du_NC2019_disorder,du_NRP2021_nonlinear,ortix_AQT2021_nonlinear}. It plays a  
fundamental role in the identification of different topological and 
magnetic states~\cite{sinha_NP2022_berry,wang_PRL2021_intrinsic}. For instance, the NL anomalous Hall 
conductivity~\cite{sodemann_PRL2015_quantum}, which determines the Hall 
response in time-reversal symmetric systems, provides information 
on the Berry curvature dipole. It also acts as a sensor for topological phase transitions of the valley-Chern type~\cite{sinha_NP2022_berry,Chakra_2Dmat2022_nonlinear}. Conversely, the intrinsic NL Hall conductivity~\cite{gao_PRL2014_field,liu_PRL2021_intrinsic} provides information on the Berry connection polarizability (BCP). Interestingly, it can sense the orientation of the N\'eel vector in parity-time reversal symmetric systems~\cite{wang_PRL2021_intrinsic}.


Most of the transport coefficients are extrinsic. In these extrinsic conductivities, the information about the electronic state of the system is entangled with the effect of disorder. 
This has motivated the search for intrinsic (scattering-independent) transport coefficients. In the linear response regime, several intrinsic Hall conductivities are known, such as the anomalous Hall~\cite{sinitsyn_IOP2007_semi,nagaosa_RMP2010_anomal,gao_N2021_layer}, spin Hall~\cite{bernevig_PRL2006_quantum,maciejko_ARCMP2011_the,konig_JPSJ2008_the}, and quantum anomalous Hall~\cite{chang_arxiv2022_quantum,liu_ARCMP2016_the} conductivities. Conversely, intrinsic responses in the NL regime are relatively less explored. Very recently, with the discovery of an intrinsic NL BCP Hall (BCPH) conductivity~\cite{gao_PRL2014_field}, this field has started to flourish. 

\begin{figure}[t]
    \centering
    \includegraphics[width = 0.98\linewidth]{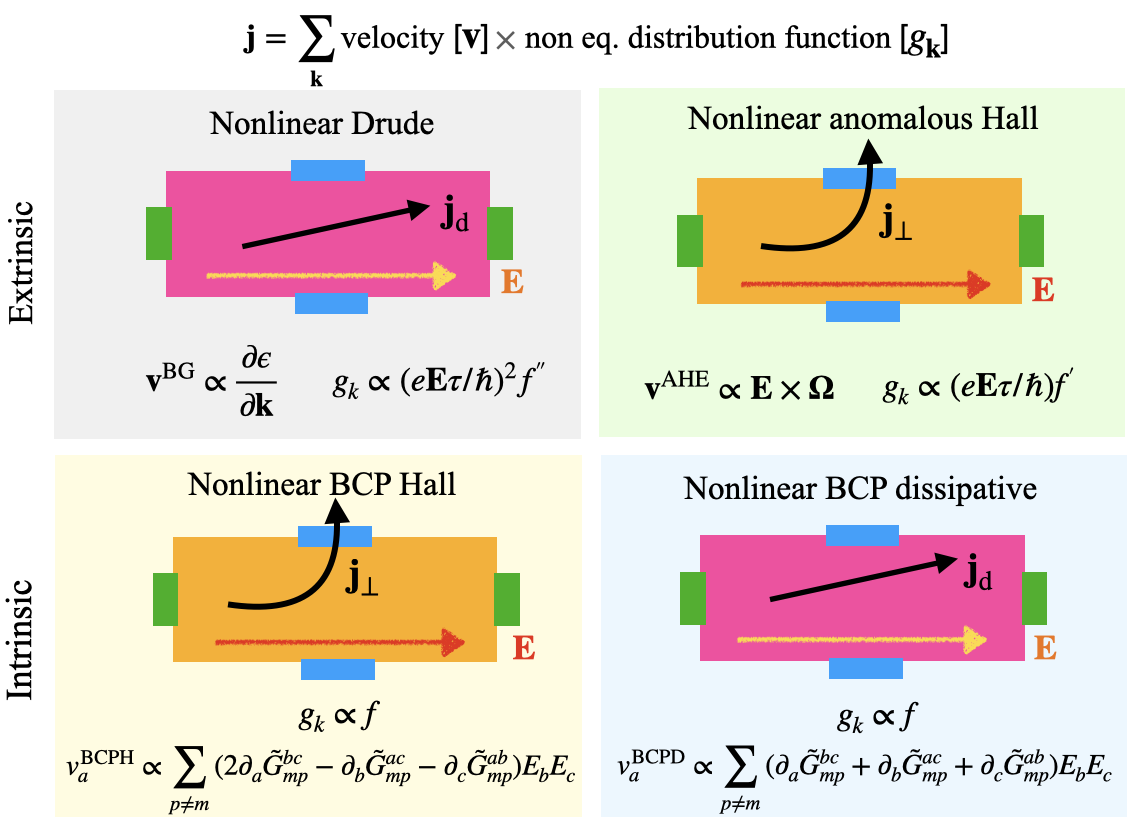}
    \caption{A schematic of all four different second-order NL transport responses in the DC limit. The two contributions in the top row depend on the scattering time. The Drude conductivity arises from the second-order correction to the distribution function and the band gradient velocity. In contrast, the anomalous Hall conductivity arises from the first-order correction to the distribution function and the anomalous Hall velocity. 
    The two intrinsic contributions are shown in the bottom row. 
    The left panel represents the nonlinear BCP Hall conductivity. 
    The right panel shows the NL BCP dissipative conductivity. 
    \label{fig_0}}
\end{figure}
%

In this paper, we predict a new {\it intrinsic} second-order NL conductivity, which gives  {rise to a dissipative current}. This new second-order NL conductivity can be expressed as
\be \label{NL_QMD}
\sigma_{a;bc}^{\rm BCPD} =\dfrac{e^3}{\hbar} \sum_{m,p, {\bm k}} \int [d{\bm k}] f_m \left[ \partial_a  {\tilde {\mathcal G}}_{mp}^{bc} +  \partial_{b} {\tilde {\mathcal G}}^{ac}_{mp} + \partial_{c} {\tilde {\mathcal G}}^{ab}_{mp} \right].
\ee%
Here, $f_m$ is the Fermi function for the $m$th band, the electronic charge is $-e$ (with $e>0$), $\epsilon_{mp}=\epsilon_m - \epsilon_p$ is the energy difference between bands, $\partial_a \equiv \partial/\partial{k_a}$, and $[d\bm k] = d^d \bm k/(2\pi)^d$ is the integration measure for a $d$-dimensional system. The quantity $\tilde {\mathcal{G}}^{bc}_{mp}={\mathcal{G}}^{bc}_{mp}/\epsilon_{mp}$ is the band normalized band-resolved quantum metric (QM) often called the Berry connection polarizability (BCP)~\cite{liu_PRL2021_intrinsic}. {The gauge invariant quantum metric ${\mathcal{G}}^{bc}_{mp}$ is the real part of the quantum geometric tensor, ${\mathcal Q}_{mp}^{bc}={\mathcal R}_{pm}^b {\mathcal R}_{mp}^c$; ${\mathcal{G}}^{bc}_{mp}=\frac{1}{2}({\mathcal R}_{pm}^b {\mathcal R}_{mp}^c+{\mathcal R}_{mp}^b {\mathcal R}_{pm}^c)$ [See Sec.~I of the Supplemental material (SM)] 
\footnote{\href{https://www.dropbox.com/s/iqkvt93vnz2lhbv/SM.pdf?dl=0}{The Supplemental material} discusses, i) Gauge invariance of the quantum geometric quantities, ii) Calculation of the density matrix, iii) NL current, iv) Quantum geometric quantities under various symmetries, iv) Other extrinsic currents and v) Tilted massive Dirac system}. Here, ${\bm {\mathcal R}}_{mp}=i\langle u_m\ket{ {\bm \nabla}_{\bm k}u_p}$ is the inter-band Berry connection with $\ket{u_p}$ being the cell periodic part of the electron wave-function. We refer to the conductivity in Eq.~\eqref{NL_QMD} as the BCP-induced dissipative (BCPD) NL conductivity. The predicted BCPD conductivity does not contribute to purely Hall response.} This can be confirmed by constructing a nonlinear purely Hall conductivity following Ref.~\cite{tsirkin_SP2022_on} as $\sigma_{\rm Hall}=\sigma_{a;bc} -\sigma_{b;ac}$ [or $\sigma_{a;bc} -\sigma_{c;ba}$]. It can be easily checked that the Hall conductivity corresponding to Eq.~\eqref{NL_QMD} vanishes identically. Its significance is manifold. The inter-band coherence effects are very strong in transverse responses such as the anomalous Hall effect \cite{Culcer_17_QKT}, but typically not in longitudinal responses. For a clean system, the only inter-band coherence effect known in longitudinal transport is Zitterbewegung \cite{Culcer_17_QKT}, which only occurs when the chemical potential lies at the Dirac point \cite{Ludwig_PRB1994}. Since, in practice, the Dirac point is always disordered, the intrinsic contribution to Zitterbewegung is, for all purposes, unobservable. Hence the BCPD conductivity can be regarded as an intrinsic quantum coherence effect in longitudinal transport in a doped system. The effect is traced to the Fermi surface and represents a quantum coherence effect in multi-band systems induced by the electric field. 



We calculate the second-order NL current within the framework of the quantum kinetic theory for the density matrix~\cite{Culcer_17_QKT,Sekine_CA2017, 
Aversa_PRB1995,khurgin_APL1995,Sipe_PRB2000,glazov_PR2014,juan_PRR2020,watanabe_PRX2021_chiral,gao_PRR2021_intrinsic}. Our quantum kinetic theory based treatment of the electric field interaction in the length gauge provides a complete picture of the NL responses, as summarized in Fig.~\ref{fig_0}. This approach includes NL electric field corrections to electron dynamics, which is missed in methods combining the first-order equation of motion of the charge carriers with the non-equilibrium distribution function \cite{lahiri_PRB2021_nonlinear}. The intrinsic conductivity defined in Eq.~\eqref{NL_QMD} vanishes in the presence of either spatial inversion symmetry ($\mathcal P$) or time-reversal symmetry ($\mathcal T$). This can be verified from the explicit form of Eq.~\eqref{NL_QMD}. In the presence of either $\mathcal T$-symmetry or $\mathcal P$-symmetry, the energy dispersion is an even function of the momentum while the band resolved quantum metric satisfies $ \mathcal{G}^{bc}_{mp}(-\bm k)=\mathcal{G}^{bc}_{mp}(\bm k)$. This combines to make Eq.~\eqref{NL_QMD} identically zero. Therefore, for the finite BCPD conductivity, both $\mathcal T$ and $\mathcal P$ must be broken. 

{\it QM induced velocity as the origin of BCPD current:---} 
In the semiclassical picture, the current is given by the product of 
the single band velocity and the corresponding non-equilibrium 
distribution function of that band. Accordingly, the NL Drude 
conductivity appears from band gradient velocity ($v^{\rm BG}_a=
\partial_a \epsilon/\hbar$) and the second order distribution function 
$f_2 = e^2 \tau^2 \partial_b \partial_c f E_b E_c$ where $E_{b/c}$ are 
the components of the electric field. The NL anomalous Hall 
conductivity arises from the electric field induced anomalous velocity~\cite{sundaram_PRB1999_wave,xiao_RMP2010_berry} 
${\bm v}^{\rm AHE}=e ({\bm E} \times {\bm \Omega})/\hbar$ and the 
first order distribution function $f_1=e \tau \partial_b f E_b$. The 
intrinsic BCPH conductivity arises from the correction of anomalous 
velocity due to electric field~\cite{gao_PRL2014_field}, ${\bm v}^{\rm BCPH}=e ({\bm E} \times 
{\bm \Omega}^{E})/\hbar$ where ${\bm \Omega}^{E}$ is the correction in the Berry curvature. 
Similarly, we attribute the BCPD conductivity to a new electric 
field-induced gauge invariant velocity called the QM-induced velocity. For the $m$-th band, it is given by 
\be \label{vel_QMD}
v_{m,a}^{\rm BCPD} =  - \frac{e^2}{\hbar}\sum_{p \neq m} \left[\partial_a \tilde{\mathcal G}_{mp}^{bc} + \partial_b \tilde{\mathcal G}_{mp}^{ac} +\partial_c \tilde{\mathcal G}_{mp}^{ab} \right] E_b E_c~.
\ee
%
In contrast to the anomalous velocity and the BCPH
velocity, the QM-induced BCPD velocity has both the longitudinal and 
the transverse components. It arises from the interband coherence effects.

{\it Quantum Kinetic theory of the second-order current:---} In the quantum kinetic theory framework, the NL current is calculated using $j_a^{(2)}=-e\sum_{m,p} v_{pm}^a \rho_{mp}^{(2)}$. Here,  
$v_{pm}^a$ and $\rho_{mp}^{(2)}$ are the velocity and the second-order 
density operator in the band basis of the unperturbed Hamiltonian, 
${\cal H}_0 \ket{u_n } = \epsilon_n  \ket{u_n}$. The velocity operator $\hat {\bm v}= (i/\hbar) [{\mathcal H}_0, {\bm r}]$. In the crystal momentum representation, it reduces to $v^a_{pm} = \hbar^{-1}\left( \partial_a 
\epsilon_p \delta_{pm} + i\mathcal{R}^a_{pm}\epsilon_{pm}\right)$. 
Here, the first term arises due to the intra-band motion of the 
electron and the second term arises from inter-band coherence ~\cite{wilczek_PRL1984_appear, culcer_PRB2005_coherent}.

The single-particle density matrix is obtained by starting from the Liouville von-Neumann equation with the Hamiltonian $\mathcal{H} = \mathcal{H}_0 + \mathcal{H}_E$.
 Here, $\mathcal{H}_E = e{\bm E \cdot\bm { r}}$ is the 
correction to Hamiltonian induced by the electric field. The NL responses of various orders are explored by expanding the density matrix perturbatively in orders of the electric field, $\rho = \rho^{(1)} + \rho^{(2)} \dots + \rho^{(N)}$, where in general we have $\rho^{(N)} \sim  |{\bf E}|^N$. The solution of the quantum kinetic equation is given by~\cite{Aversa_PRB1995}
\be \label{rho_N}
i\hbar \tilde \rho^{(N+1)}(t) = e \int_{-\infty}^{t} dt^\prime e^{\frac{i}{\hbar} {\mathcal H}_0 t^\prime}{\bm E}(t^\prime) \cdot \left[{\bm r}, \rho^{(N)}(t')\right]e^{-\frac{i}{\hbar} {\mathcal H}_0 t^\prime}~.
\ee
In the following, we consider ${\bm E}(t)={\bm E} e^{-i \omega t} e^{- \eta |t|}$ (adiabatic switching approach) and finally put $\omega=0$ for the DC transport results. The tilde represents the density matrix in the interaction picture. We assume the zeroth order (or equilibrium) density matrix to be $\rho^{(0)}_{mp} = f_m\delta_{mp}$, where $f_m = [1+ e^{\beta(\epsilon_m-\mu)}]^{-1}$ is the Fermi-Dirac distribution with $\beta  = 1/(k_BT)$, $k_B$ is the Boltzmann constant, $T$ is the absolute temperature and $\mu$ is the chemical potential. For convenience, we express the second-order density matrix as a sum of four parts~\cite{Aversa_PRB1995,watanabe_PRX2021_chiral}: two in the diagonal $\rho_{mm}^{\rm dd}$,  $\rho_{mm}^{\rm do}$ and two in the off-diagonal $\rho_{mp}^{\rm od}$ and $\rho_{mp}^{\rm oo}$ sector. Here, the first superscript indicates the diagonal ($\rm d$) or off-diagonal ($\rm o$) nature of the second-order density matrix. The second superscript indicates the corresponding contribution from the first-order density matrix, i.e. $\rho^{(1)}$ inside the commutator of the right-hand side of Eq.~\eqref{rho_N} (See Sec.~II of the SM~\cite{Note1}).

The second-order current can be separated into three parts: $j^{(2)}_a= j^{(2)}_a (\tau^0) +  j^{(2)}_a (\tau^1) + j^{(2)}_a (\tau^2)$. The element $\rho_{mm}^{\rm dd}$ does not contribute to any intrinsic current, while all the other elements contribute. We denote the intrinsic part stemming from $\rho_{mm}^{\rm do}$, $\rho_{mp}^{\rm od}$ and $\rho_{mp}^{\rm oo}$ as $j^{\rm int, do}_a$, $j^{\rm int, od}_a$, and $j^{\rm int, oo}_a$, respectively. 
These three provide the complete set of intrinsic contributions to the current  $j^{\rm int}_{a} = j^{\rm int, do}_a + j^{\rm int, od}_a + j^{\rm int, oo}_a$. The corresponding intrinsic conductivity is, 
\be \label{total_int}
\sigma^{\rm int}_{a;bc} = -\dfrac{e^3}{\hbar} \sum_{m,p, {\bm k}} f_m \left[ \partial_a  \tilde {\mathcal G}_{mp}^{bc} - 2 \left( \partial_{b} \tilde {\mathcal{G}}^{ac}_{mp} + \partial_{c} \tilde {\mathcal{G}}^{ab}_{mp}\right) \right]. 
\ee
For calculation details, see Sec.~III of the SM~\cite{Note1}. {This is the main result of our paper and the  physically relevant nonlinear intrinsic conductivity.}
Comparing this intrinsic contribution to the existing semiclassical results for the intrinsic conductivity~\cite{gao_PRL2014_field},  {we find that it naturally separates into (dissipationless) Hall and dissipative components~\footnote{Following Ref.~[\onlinecite{tsirkin_SP2022_on}], we note that the separation of any nonlinear conductivity into dissipative and non-dissipative part is not unique and there are other possibilities as well. However, the fully symmetrized part of the conductivity in Eq.~(1) is a reasonable choice for the dissipative part, and the remaining component is the non-dissipative or Hall part. In our case, this choice also aligns with the existing literature and the nondissipative conductivity contribution in Eq.~(4) is identical to that obtained via Boltzmann transport approach in Ref.~[11].}} 
as $\sigma^{\rm int}_{a;bc}=\sigma^{\rm BCPH}_{a;bc}+\sigma^{\rm BCPD}_{a;bc}$. Here, the BCPH part represents the purely Hall response and is given by~\cite{gao_PRL2014_field,wang_PRL2021_intrinsic,liu_PRL2021_intrinsic}
\be \label{NL_BCP}
\sigma_{a;bc}^{\rm BCPH} =-\dfrac{e^3}{\hbar}\sum_{m,p, {\bm k}} f_m \left[2 \partial_{a} {\tilde {\mathcal G }}^{bc}_{mp}-\left(\partial_{b} {\tilde {\mathcal G}}^{ac}_{mp} +\partial_{c} {\tilde {\mathcal{G}}}^{ab}_{mp}\right)\right],
\ee 
and the other part, which represents the dissipative response, is given in Eq.~\eqref{NL_QMD}.
We would like to mention that  {$\tilde {\mathcal G}$ used in this paper is half of what has been denoted as $G$ in Ref.~\cite{gao_PRL2014_field}}. Furthermore, to compare our results with Ref.~\cite{gao_PRL2014_field}, we symmetrize their results~\cite{oiwa_JPSJ2022_sys} in the field (last two) indices.  {We emphasize that although the purely Hall conductivity in Eq.~\eqref{NL_BCP} and the BCPD contributions require the same fundamental symmetry restriction, the constraints of the crystalline symmetries are different. Therefore even if the purely Hall current vanishes, the contribution from Eq.~\eqref{NL_QMD} can still be finite.}

{\it Tilted massive Dirac system:---}  We choose the tilted Dirac system as it offers several insights into different NL BCPD conductivity contributions while being analytically tractable. The Hamiltonian we consider is given by~\cite{lahiri_PRB2021_nonlinear},  
\be \label{ham}
{\mathcal H} = v_F  (k_x \sigma_y -  k_y\sigma_x) +   v_{t} k_y \sigma_0 + \Delta\sigma_z~.
\ee
Here, $v_F$ is the Fermi velocity, $\sigma_i$'s are 
the Pauli matrices representing the sub-lattice 
degree of freedom, $\Delta$ is the gap in the system 
and the $v_t$ term introduces tilt along the $k_y$-axis. 
This Hamiltonian breaks both $\mathcal{T}$- and 
$\mathcal{P}$-symmetry. 
The dispersion for this two band model is given by $\epsilon_{\pm} = v_t k_y \pm \epsilon_0$, where $\epsilon_0 = \left(v_F^2 k^2 + \Delta^2\right)^{1/2}$ with $k = \left(k_x^2 + k_y^2\right)^{1/2}$. 
%
%
%
The various elements of the quantum metric for this model Hamiltonian is calculated to be
\be 
\begin{pmatrix}
{\mathcal G}_{cv}^{xx} &  {\mathcal G}_{cv}^{xy} \\
{\mathcal G}_{cv}^{yx} & {\mathcal G}_{cv}^{yy}
\end{pmatrix}
=\frac{v_F^2}{4 \epsilon_0^4} 
\begin{pmatrix}
k_y^2 v_F^2 + \Delta^2 &  - v_F^2 k_x k_y\\
- v_F^2 k_x k_y & k_x^2 v_F^2 + \Delta^2
\end{pmatrix}~.
\ee%
The quantum metric for 
this model is independent of the tilt velocity as expected. In 
contrast to the Berry curvature, the gap parameter 
$\Delta$ is not essential to have a finite quantum 
metric. In the context of 2D hexagonal Dirac systems such as graphene, 
the gap opening is associated with inversion symmetry breaking. For graphene, the inversion symmetry breaking is physically associated with the $A$ and the $B$ sublattice having different onsite potential induced by the substrate.
This highlights that the quantum metric can 
be finite even in the presence of both the ${\mathcal 
P}$ and ${\mathcal T}$ symmetries. 

\begin{figure}[t]
    \centering
    \includegraphics[width = 0.9\linewidth]{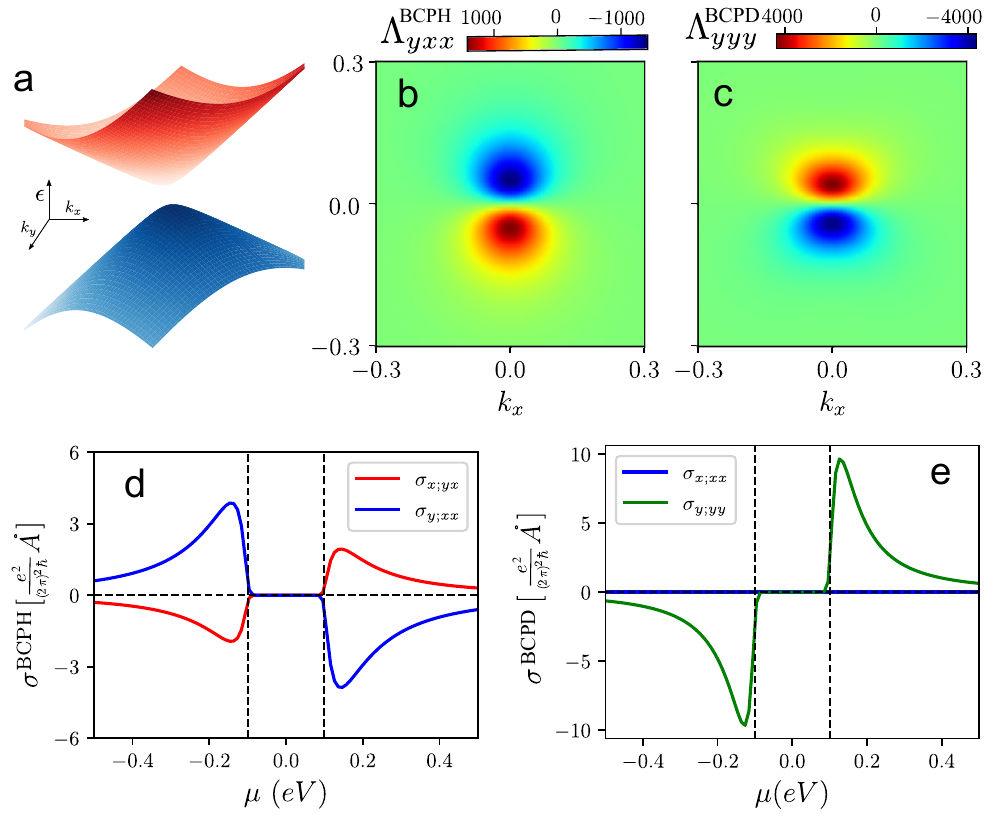}
    \caption{a) Schematic of the dispersion of the tilted massive Dirac model. b), c) The momentum space distribution of the BCPH and BCPD components of the dipoles.  {They are in units of ${\rm eV}^{-1}$\AA$^{-3}$.} d) Variation of the non-zero BCPH contributions with chemical potential, $\mu$. e) The BCP-induced NL  {dissipative} conductivities. The various parameters for the Hamiltonian are chosen to be $\Delta = 0.1$eV, $v_t = 0.1$eV~\AA, and $v_F = 1$eV ~\AA. We have considered temperature $T= 50$ K.  }
    \label{fig2}
\end{figure}

We present the distribution of the band geometric quantities in the momentum space in Fig.~\ref{fig2}. Panel (a) shows a schematic of the dispersion of the tilted massive Dirac model. In panel (b), we have shown the BCPH dipole component $\Lambda_{yxx}^{\rm BCPH}$ for the valence band. The BCPH dipole (for band $m$) is defined as~\cite{liu_PRL2021_intrinsic}  {
\be 
\Lambda_{abc, m}^{\rm BCPH} = \sum_p\left[2 \partial_{a} {\tilde {\mathcal G}}^{bc}_{mp} - \partial_{b} {\tilde {\mathcal G}}^{ac}_{mp} - \partial_{c} {\tilde {\mathcal G}}^{ab}_{mp} \right] f_m~.
\ee}
We note that the component of the BCP dipole  
show a dipole-like behavior in the momentum space 
distribution [see Fig.~\ref{fig2}(b)]. Similarly, for 
the BCPD conductivity, we have defined the quantum 
metric-dependent BCPD  dipole (for band $m$) as 
\be \label{Lambda_QMD}
\Lambda_{abc, m}^{\rm BCPD} =\sum_{p} \left(\partial_a {\tilde {\mathcal G}}^{bc}_{mp} + \partial_b {\tilde {\mathcal G}}^{ac}_{mp}+ \partial_c {\tilde {\mathcal G}}^{ab}_{mp}\right)  f_m~.
\ee%
We have plotted the $\Lambda_{yyy}^{\rm BCPD}$ component in Fig.~\ref{fig2}(c), and it shows dipolar behavior.

We have calculated the intrinsic NL 
transport coefficients for this model Hamiltonian, in the small tilt 
limit $v_t/v_F \ll 1$.
Assuming $\mu> \Delta$ and defining  $r=\Delta/\mu$ for brevity, we obtain for the conduction band (See Sec. V of SM~\cite{Note1} for details), 
\bea
\sigma_{y;yy}^{\rm BCPD} &=& \frac{15 e^3  v_t }{128 \pi \hbar \mu^2}  \left[1 + 2r^2   - 3 r^4 \right],~~~~
\\
\sigma_{y;xx}^{\rm BCPH} &=& - \frac{ e^3  v_t }{8 \pi \hbar \mu^2}  \left[1 -  r^2 \right]. 
\eea
Both the BCPH and the BCPD conductivities can be finite even in the absence of a gap, i.e., in the limit $\Delta \to 0$ or $r \to 0$ with finite $\mu$. This can be 
understood from the fact that in contrast to the Berry 
curvature, the quantum metric can be finite even in the presence of both of 
the $\cal T$ and the $\cal P$ symmetries. 
However, both of these quantities depend on the tilt velocity  
and vanish if $v_t \to 0$. In Fig.~\ref{fig2}, we 
have shown both the intrinsic conductivities. Both these  
conductivities change their sign when going from the valence band 
to the conduction band. 
Since the BCPH and BCPD conductivity are a Fermi surface 
effects, it is expected that it will vanish in the band gap. 

{\it ${\mathcal P}{\mathcal T}$ symmetric CuMnAs:---}
CuMnAs has anti-ferromagnetic ordering, with opposite spins lying on a bipartite lattice. Such an arrangement breaks the $\mathcal P$ as well as the $\mathcal T$ symmetry locally. However, the combined ${\mathcal P}{\mathcal T}$ symmetry is preserved by the exchange of the sub-lattices with the flip of oppositely aligned spins~\cite{watanabe_PRX2021_chiral}. The model Hamiltonian for CuMnAs is given by
\be \label{H_cumnas}
\mathcal{H}(\bm k) = 
\begin{pmatrix}
\epsilon_0(\bm k) + {\bm h_{\rm A}( k)\cdot \pmb{\sigma}} & V_{\rm AB}(\bm k) \\
V_{\rm AB}(\bm k) & \epsilon_0(\bm k) + {\bm h_{\rm B}( k)\cdot \pmb{\sigma}} 
\end{pmatrix}~.
\ee 
Here, $\epsilon_0({\bm k}) = -t (\cos k_x + \cos k_y)$ and $V_{\rm AB}({\bm k}) = -2 \tilde t \cos({k_x}/{2}) \cos({k_y}/{2})$ where $t$ and $\tilde t$ denotes hopping between orbitals of the same and different sub-lattices, respectively. The sub-lattice dependent spin-orbit coupling and the magnetization field are included in ${\bm h}_{\rm B}({\bm k})=-{\bm h}_{\rm A}({\bm k})$, where ${\bm h}_{\rm A}({\bm k}) = \{h_{\rm AF}^x - \alpha_{\rm R} \sin k_y + \alpha_{\rm D} \sin k_y, h_{\rm AF}^y + \alpha_{\rm R} \sin k_x + \alpha_{\rm D} \sin k_x, h_{\rm AF}^z\}$. Here, $\alpha_{\rm R}$ and $\alpha_{\rm D}$ represent the Rashba and the Dresselhaus spin-orbit coupling, respectively. 

\begin{figure}[t!]
    \centering
    \includegraphics[width=\linewidth]{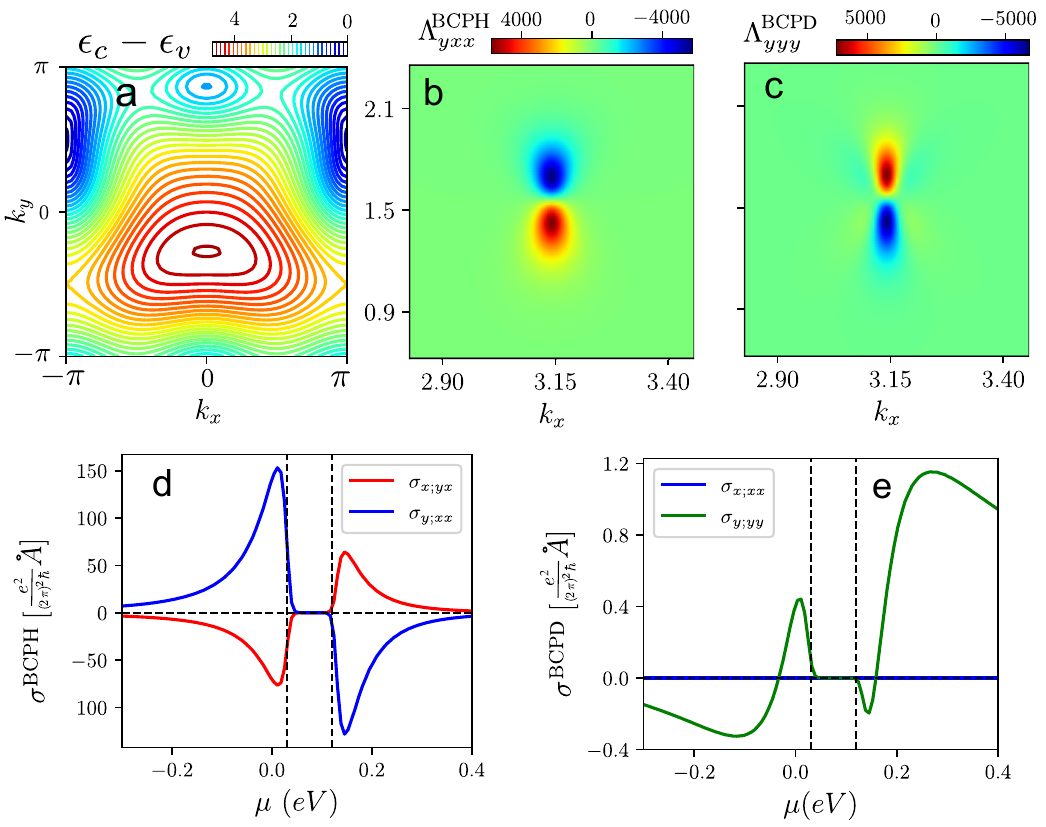}
    \caption{ a) The energy gap between the conduction and the valence band in units of eV. Note the gapped Dirac points near $(k_x,k_y)=(\pm 1, 0.5)\pi$ and $(k_x,k_y)=(0, 0.8)\pi$. Near $(\pm 1, 0.5)\pi$, the BCP Hall dipole is shown in b), and the BCP longitudinal dipole is shown in c).  {They are in units of ${\rm eV}^{-1}$\AA$^{-3}$.} d) The chemical potential dependence of the NL Hall conductivity in which the contributions are induced by the BCP Hall dipole. e) The chemical potential dependence of the longitudinal nonlinear conductivity induced by the BCP longitudinal dipole. We have used the Hamiltonian parameters $t=0.08$ eV and $\tilde t=1$ eV. The other parameters are $\alpha_{\rm R}=0.8, \alpha_{\rm D}=0$ and ${\bm h}_{\rm AF}=(0, 0, 0.85)$ eV.  For the conductivity calculation we have considered temperature $T= 50$ K.}
    \label{fig_4}
\end{figure}

Depending on the various parameters of the Hamiltonian, one can have an
insulating state, a gapless state, or a gapped Dirac state as the ground state. Here, we work 
with the gapped Dirac phase, where two gapped Dirac points appear near 
the zone boundary at the extremes of the $k_x$-axis in the positive half 
of the $k_y$-axis as shown in Fig.~\ref{fig_4}(a). We have highlighted 
the corresponding 
BCPH dipole in Fig.~\ref{fig_4}(b) and BCPD dipole in Fig.~\ref{fig_4}(c), respectively, 
in vicinity of $(k_x,k_y) = (1, 0.5) \pi $. 
To demonstrate the intrinsic Hall and longitudinal conductivity, we show the $\mu$ dependence of the BCPH 
conductivity along with the NL BCPD conductivity in 
Fig.~\ref{fig_4}. We find that the BCPH conductivity $\sigma_{\rm H}=(\sigma_{y;xx}-\sigma_{x;yx})$ is non-zero in this system. More importantly, the NL BCPD conductivity, induced by the QM contribution, is also non-zero.

{\it Discussion:---} The recent interest in  intrinsic contributions to the second-order NL conductivities was  triggered by the prediction of intrinsic NL anomalous Hall effect in Ref.~[\onlinecite{gao_PRL2014_field}] using the semiclassical wave-packet formalism. Since then, this problem has been approached using different methods. Unfortunately, different approaches lead to slightly different results. For instance, using the velocity gauge approach, a Fermi sea contribution in the NL conductivity was reported in Ref.~\cite{oiwa_JPSJ2022_sys}. A noncyclic longitudinal conductivity has been obtained in Ref.~\cite{oiwa_JPSJ2022_sys,kaplan_arxiv2022_unif}, which is attributed to the mixed axial-gravitational anomaly~\cite{holder_arxiv2021_mixed}. An in-gap NL Hall conductivity has been proposed in Ref.~\cite{kaplan_arxiv2022_general}. In the length gauge approach, we find that Ref.~\cite{nandy_PRB2019_symmetry} 
 and Ref.~\cite{wang_arxiv2022_an} also obtained an intrinsic NL conductivity. An intrinsic scattering time-independent photogalvanic response was reported in Ref.~\cite{gao_PRR2021_intrinsic} and in Ref.~\cite{watanabe_PRX2021_chiral}.
 
{In our calculation, we find that the choice of relaxation time is crucial in the nonlinear regime. If we consider $\tau$ instead of $\tau/2$ for the second order density matrix, then $\rho^{\rm int, do} \to 2 \rho^{\rm int, do}$, $\rho^{\rm int, od} \to \frac{1}{2} \rho^{\rm int, od}$ and $\rho^{\rm oo}$ remains unchanged. Although this reproduces the purely Hall contribution of Ref.~\cite{gao_PRL2014_field}, it inevitably it leads to an in-gap dissipative current of the form $j_{\rm gap}=\frac{e^3}{\hbar}\sum_{m,n} (\partial_a  {\mathcal G }^{bc}_{mn}/\omega_{mn} )f_n E_bE_c$ which is unphysical.} This has also been highlighted in Ref.~\cite{gao_PRR2021_intrinsic,oiwa_JPSJ2022_sys,kaplan_arxiv2022_unifying}. Using the adiabatic perturbation theory approach within the density matrix framework, we find that the intrinsic Hall response of the systems is only dictated by the BCP contribution predicted by Gao et al. ~\cite{gao_PRL2014_field}. The additional NL conductivity we obtained is cyclic in all the spatial indices. We did not obtain any in-gap conductivity (neither Hall nor longitudinal). 

{\it Conclusion:---} To conclude, we unravel the physics of interband coherence due to electric field in intrinsic NL transport using the quantum kinetic theory framework. In addition to providing the quantum 
kinetic theory of recently discovered BCP-induced NL Hall conductivity, here we predict a new intrinsic NL conductivity. Remarkably, this conductivity is dissipative and gives rise to an intrinsic longitudinal current which we termed BCPD conductivity. This newly discovered current brings a new term to the intrinsic NL effect, and, more importantly, it is the only example of longitudinal transport arising from quantum coherence effects in doped systems. 

This newly discovered conductivity broadens our present 
understanding of NL transport phenomena. 
Following our electronic transport calculations, thermal and thermoelectric~\cite{mandal_PRB2020_magnus,das_PRR2020_thermal,das_PRB2021_intrinsic} intrinsic transport may also display interesting NL effects. Non-trivial physics may additionally emerge in the presence of magnetic fields with previously unexplored intrinsic magneto-transport phenomena~\cite{gao_PRL2014_field, Gao_OMS_2015}. 

\acknowledgments

We thank Leonid Golub for a series of enlightening discussions. K. D. thanks Daniel Kaplan for insightful discussions. K. D. is supported by the Weizmann Institute of Science and the Koshland Foundation. DC is supported by the Australian Research Council Centre of Excellence in Future Low-Energy Electronics Technologies (project number CE170100039). We acknowledge the Science and Engineering Research Board (SERB, for the MTR/2019/001520 project), and the Department of Science and Technology (DST, for the project DST/NM/TUE/QM-6/2019(G)-IIT Kanpur) of the Government of India, for financial support. We thank Harsh Varshney for carefully reading our manuscript and checking the calculations. We thank one of the anonymous referees.


\bibliography{ref.bib}

\end{document}